\documentclass[english,usenatbib]{article}
\usepackage{babel,amssymb,graphicx}
\usepackage{epsfig}
\usepackage{natbib}

\begin{document}
\title{Absence of a Periodic Component in
       Quasar $z$-Distribution}
\author
{S.V.~Repin$^{1}$, \quad B.V.~Komberg$^{2}$, \quad
V.N.~Lukash$^{2}$
\\ \\
$^{1}$  Space Research Institute of RAS, Moscow, \\
$^{2}$  Astro-Space Center of Levedev Physical Institute, Moscow}


\maketitle

\begin{abstract}
Since the discovery of quasars in papers often appeared and appear
the asser\-tions that the redshift quasar distribution includes a
periodic component with the period $\Delta z = 0.063$ or 0.11. A
statement of such kind, if it is correct, may manifest the
existence of a far order in quasar distribution in cosmological
time, that might lead to a fundamental revision all the
cosmological paradigm. In the present time there is a unique
opportunity to check this statement with a high precision, using
the rich statictics of 2dF and SDSS catalogues ($\sim$ 85000
quasars).  Our analysis indicates that the periodic component in
distribution of quasar redshifts is absent at high confidence
level.

\end{abstract}

\section{Introduction}

     As early as the first hundred galaxies with active nuclei and
quasars have been discovered, the attempts to reveal the
periodicity in their redshift distribution have been made. For
example, the presence of peaks at $z_* = 0.061\cdot n$, where  $n$
is the integer, for the distribution of 73 objects with
non-thermal optical continuum and $z < 0.6$ was mentioned in
Burbidges papers \citep{burbidge_1957,burbidge_1968}. They used
this fact to confirm their hypothesis concerning the
non-cosmological origin of the lines redshift in active galaxy and
quasar spectra. However, the other interpretations are also
possible. Particularly, some authors discussed the effect of the
influence of occurence of several strong emission lines typical
for quasars (Mg II, 2800 \AA; C III, 1900 \AA; C IV, 1550 \AA;
Ly$_{\alpha}$, 1216~\AA), in the range of spectral observations
($\lambda > 3300$ \AA), which might emulate the ''humps'' in
$N_q(z)$ distribution (see, for example \cite{karitskaya_1970}).

     In the following years in a numbre of papers
(e.g. \cite{jaakkola_1971,tifft_1976,tifft_1989,tifft_1993,
tifft_1996, bell_2003,narlikar_1993,bell_2004a}) the authors
reported about the observed quantization of the redshifts in
spectra of near S-galaxies, which satisfies the ''Tifft series'':
\begin{equation}
  \Delta v_r =  2^{-(D + B/9)},
\end{equation}
where $v_r$ is the radial velocity expressed in light velocity
units,  $D$ and $B$ are the integers.

     In \cite{khod_1979,khod_1990} papers
the cyclical changes of statisticslly brightest quasars were
revealed in V filter, using the argument
\begin{equation}
  x \equiv \ln\left(1+z\right)
  \label{x_variable}
\end{equation}
with  $\Delta x \approx 0.19$ period. In \cite{khod_1988} paper
the dependence of a numbre of powerful radiopulsars has been
plotted against $x$ variable. In the centimeter wavelength region
the cyclic changes were revealed with periods  0.12, 0.19 and
0.38. It is interesting that much later in
\cite{ryabnikov_2001,ryabnikov_2001a} papers, the similar result
was obtained after the analysis of the redshist distribution of
$\sim$ 800 absorption lines in spectra of bright AGN with $0.10 <
z < 3.7$. The authors mentioned that a separate analysis of
$N_{abs}(z)$ distribution in different celestial hemispheres
indicates that the phase of periodicity is conserved. They
interpreted such unexpected result as a sign of existence of an
oscillatory regime in the Universe expansion and then as a
presence of a large scale cellular structure in the distribution
of the absorption systems in quasar spectra. In
\cite{ryabnikov_2010} paper the presence of the periodic structure
is considered on the basis of the absorbtion lines.

     In \cite{karlsson_1971,karlsson_1974}
papers  the peaks in $N_q(z)$ distribution with the step $\Delta x
\approx 0.19$ were considered on the basis of 574 quasar data. In
papers \cite{burbidge_2001} and \cite{bell_2006} on the basis of a
gross sample the peaks in $N_q(z)$ distribution were mentioned at
$z_* = 0.062$, 0.3, 0.6, 0.96, 1.41, 1.96, 2.63, 3.45. These peak
values satisfied the ''Bell series'', \cite{bell_2002}:
\begin{equation}
  z_* = 0.062 \cdot \left( 10N - M\right),
\end{equation}
where $N$ and $M$ take on integer values (see the corresponding
table in \cite{bell_2002}). The initial series redshift coincides
with the first term of the main Tifft series with $D=4$, $B=0$,
that corresponds to the velocity $\approx$~18600~km/s.

     It is obvious that in order to confirm such a non-standard conclusion
concerning the distinctive features in redshift quasar
distribution it is necessary to analyse a much wider statistical
information, which is included in the catalogues 2dF \citep{TwoDF}
and SDSS \citep{SDSS7}. And the papers with analysis of such kind
have really appeared. Thus, for example, in \cite{tang_2005} paper
there was reported the result of analysis of 290 quasars sample
(the same sample as used in Karlsson and Burbidge papers) and the
periodicity with the step $\Delta x = 0.081$ has been confirmed at
$3\,\sigma$ level. However, the analysis of larger samples (22497
quasars in 2dF and 46420 in SDSS-5), presented in the same paper
\cite{tang_2005}, did not reveal any periodicity. From this they
drew a conclusion that the point is in non-homogeneity of small
samples and the selection effects, which the small samples are
exposed. So, the problem, seemingly, can be abandoned,

     However, there is a numbre of papers where the authors,
analyzing the large catalogues, find the arguments for the
existence of the periodicity in $N_q(z)$ distribution. For
example, in \cite{bell_2006} paper, according to SDSS data 6 peaks
in the power spectrum were detected with the step $\Delta z =
0.65$ at $z_* = 1.2$, 1.8, 2.4, 3.1 and 3.7. And in papers
\cite{hurtnett_2007a}, \cite{hurtnett_2007b} on the basis of 2dF
and SDSS catalogues the authors reported the presence of the
''humps'' through $\Delta z = 0.0102, 0.0246, 0.0448$, that
assuming that $H_0 = 72$~km/c~Mpc corresponds to the cells of 44,
102 and 176~Mpc. Except that, the distribution $N_q(z)$ is
represented in the form $z_* = 0.062\cdot n$, where
$n=3,4,5,6,10,20$ \citep{hurtnett_2007c}. Nevertheless, in
\cite{tang_2005} paper the arguments have been expressed for the
fact that SDSS catalogue may include a periodic component because
of the selection effects.

     However, it is clear that the analysis of different
quasar samples leads the authors to unlike conclusions, concerning
the existence or non-existence of a far order in $N_q(x)$
distribution. At present there is a unique possibility to check
with high precision the hypothesis about the possible periodicity,
using the rich statistics of 2dF and SDSS catalogues ($\sim$ 85
thousand quasars). So, it is worth to consider this problem more
accurately, and this is the goal of the paper.

\section{Observational data and periodicity extraction methods}

\begin{figure}
  \centerline{
  \includegraphics[width=\columnwidth]{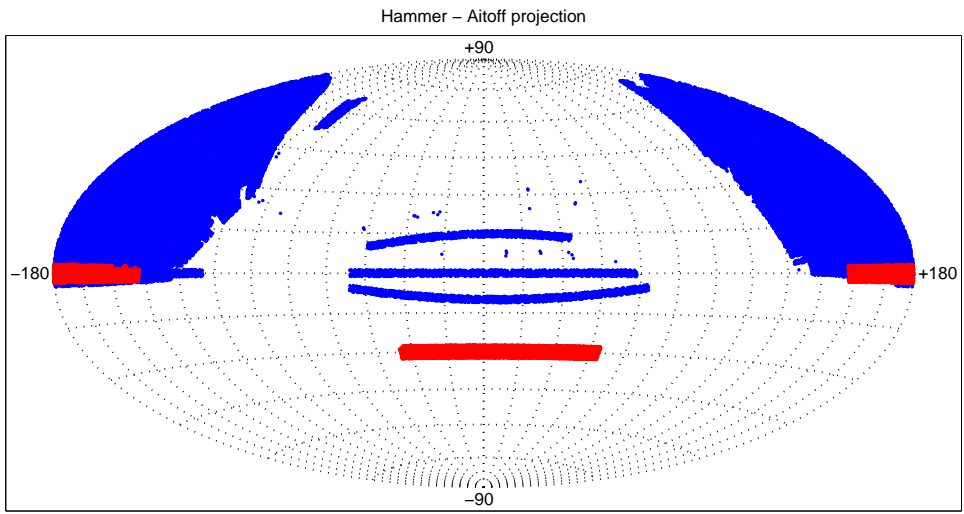}
  }
  \caption{Covering the celestial sphere by 2dF (dark grey) and
           SDSS (light grey) catalogues.}
  \label{covering_2_7}
\end{figure}

     In the current paper we used two catalogues: 2dF \citep{TwoDF}
(22272 objects) and SDSS \citep{SDSS7} (63255 objects, release 7).
Covering the celestial sphere by these catalogues is shown in
Fig.~\ref{covering_2_7}. As one can see, these regions are
overlapped and the density of the objects in 2dF catalogue is as
much as one order larger than in SDSS. In last (seventh) release
of SDSS catalogue some ''gaps'' have been filled with respect to
the previous one, and as a result of it the sample become more
homogeneous.

    The periodicity criterion is a function $K(T)$ of the data and
a trial period  $T$, which takes on ''large'' values when the data
includes the periodic component with period $T$, otherwise its
values are ''small''.

    To investigate the periodicity we used four different
criteria. The first of them and, apparently, the most familiar one
is the Rayleigh criterion:
\begin{equation}
   K_1(T) = \frac{2}{N}
             \left[
              \left(\sum\limits_{i=1}^N \sin\frac{2\pi x_i}{T}
              \right)^2 +
              \left(\sum\limits_{i=1}^N \cos\frac{2\pi x_i}{T}
              \right)^2
             \right],
   \label{Rayleigh_criterion}
\end{equation}
where $x_i$ are the values of the variable $x$ from
(\ref{x_variable}) for each quasar with redshift $z$, $N$~--~total
amount of objects, and $T$ -- trial period in the units of $x$.
Three other criteria analyse the structure of assumed periodic
component (for variable stars it is called a light curve). For
that purpose we subdivide the assumed trial period $T$ in $m$
parts (each of $T/m$ length), calculate the phase of each quasar
and add the unity in the appropriate $m$-th part of the period. As
a result we obtain a histogram, which indicates the numbre of
quasars that drop in each of $m$ parts of assumed period $T$.
Three criteria for that histogram analysis are the variety of the
epoch superposition method and can be written as:
\begin{equation}
   K_2(T) = \frac{m}{N}
             \sum\limits_{j=1}^m
              \left( n_j - \frac{N}{m} \right)^2,
   \label{Second_criterion}
\end{equation}
\begin{equation}
   K_3(T) = 1 - \frac{\min n_j}{\max n_j},
   \label{Third_criterion}
\end{equation}
\begin{equation}
   K_4(T) = \frac{m}{N}
            \left[\sum\limits_{j=1}^{m-1}
              \left( n_{j+1} - n_j\right)^2 +
              \left( n_1 - n_m\right)^2
            \right],
   \label{Fourth_criterion}
\end{equation}
where $m$ is the number of parts (bins, intervals) in which the
trial period is subdivided and  $n_j \;\; (j=1,2,\dots,m)$ -- the
numbre of quasars which drop in the appropriate $j$-th part of a
trial period. If the periodic component in the data is absent,
then all $m$ bins should have approximately the same values. On
the contrary, if the data contain a periodic component they should
strongly differ. All criteria have different sensitivity and
reveal different aspects of periodicity, therefore it is not
unreasonable to use all of them for a more reliable detection of
the periodicity.

    If we use the stochastic data, the mean criteria values are:
$MK_1 = 2$, $MK_2 = 9$, $MK_4 = 20$, that corresponds to the
accepted value $m = 10$, and the dispersions are, respectively,
$DK_1 = 4$, $DK_2 = 18$, $DK_4 = 120$. It means that if the
criterion $K_1$ takes on the value $K_1=7$, then it exeeds the
random signal level by $2.5\,\sigma$. To calculate the theoretical
values of $MK_3$ and $DK_3$ we should make use the formulas from
\cite{Gurin_1988a}. Statistical properties of the criteria are
considered in~\cite{gurin_1992} paper.

\begin{figure}
  \centerline{
  \includegraphics[width=\columnwidth]{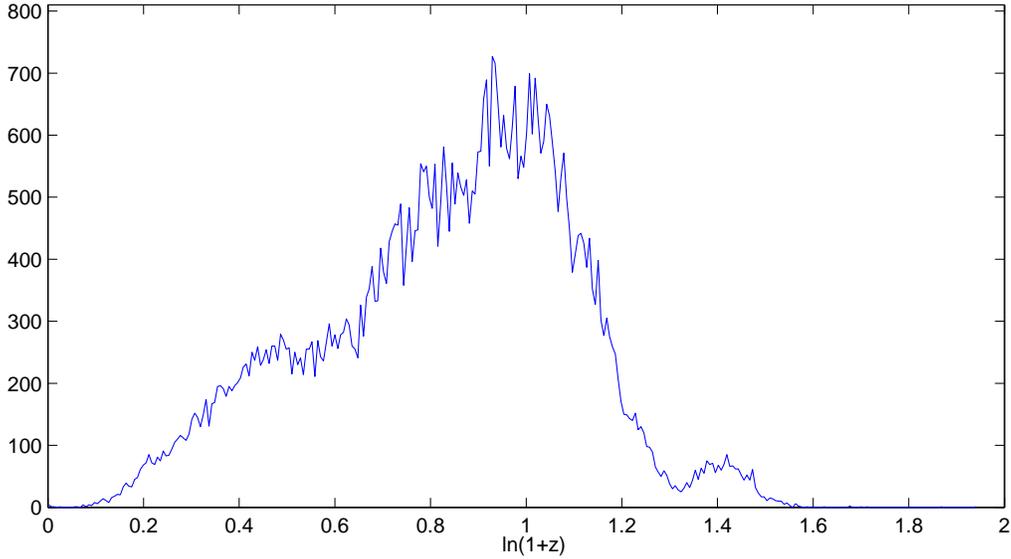}
  }
  \caption{Quasar distribution over redshift in the SDSS catalogue.
           The absciss axis corresponds to the quantity
           $x\equiv\ln(1+z)$ from (\ref{x_variable}).
           The ordinate axis shows the numbre of objects
           in the $x$-direction interval of 0.004 width.}
  \label{QSODistrSDSS7}
\end{figure}

\section{Analysis of quasar distribution over redshift}

\begin{figure}
  \centerline{
  \includegraphics[width=\columnwidth]{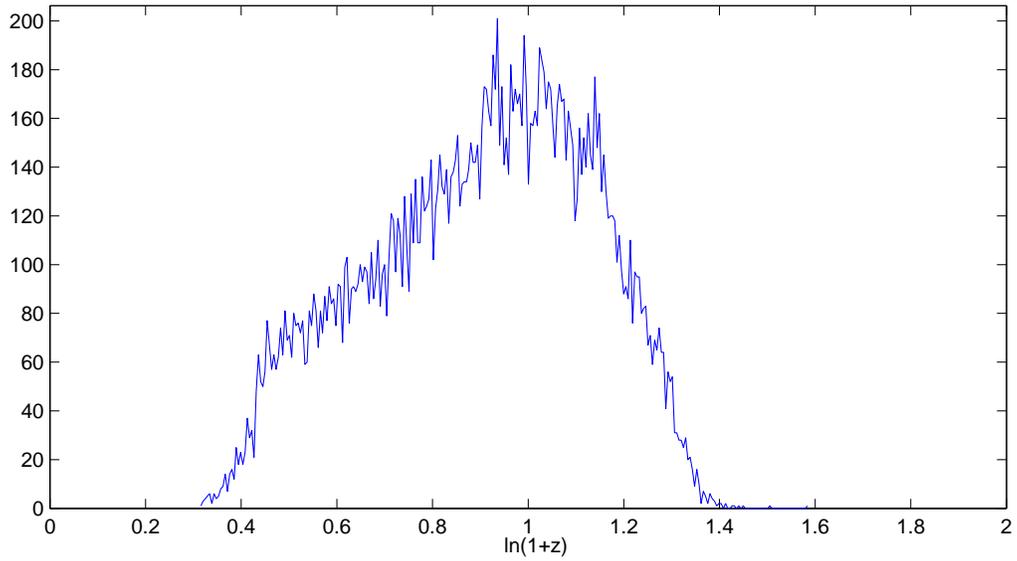}
  }
  \caption{Quasar distribution over redshift in the 2dF catalogue.
           The ordinate axis shows the numbre of objects
           in the $x$-direction interval of 0.0046 width.}
  \label{QSODistr2dF}
\end{figure}

     The quasar distribution over redshift $z$ in two catalogues shown
in Fig.~\ref{QSODistrSDSS7} and~\ref{QSODistr2dF}. Analysing the
plots one can draw a conclusion that the SDSS catalogue is more
representative in the region of large and small $x$ ($x<0.4$ and
$x>1.3$). Four large maxima in the SDSS quasar distribution near
$x_* \approx 0.5, 0.8, 1$ è 1.4 can be explained by the selection
effect (the redshift $z$ is detected using four spectral lines)
and are not related to a periodicity. In general, we may observe
in the plots some oscillations with smaller periods, which might
appear a weak periodic component after a detailed analysis.

\begin{figure*}
  \centerline{
  \includegraphics[width=\textwidth]{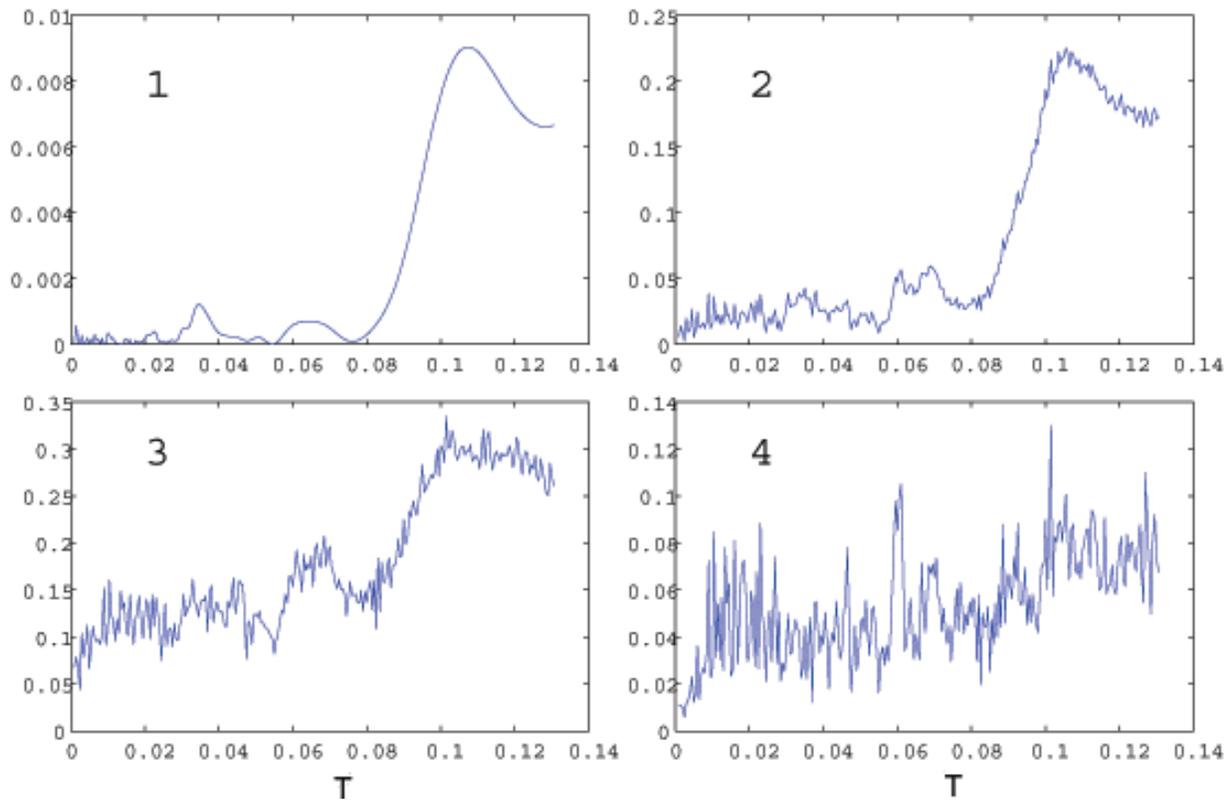}
  }
  \caption{Spectrum $K(T)$ from
           (\ref{Rayleigh_criterion}) -- (\ref{Fourth_criterion})
           of quasar distribution $N_q(T)$ according to the SDSS
           catalogue. The absciss corresponds to the trial period $T$.
           The criterion numbre marked in each panel.}
  \label{SDSSSpectrum1}
\end{figure*}

\begin{figure}
  \centerline{
  \includegraphics[width=10cm]{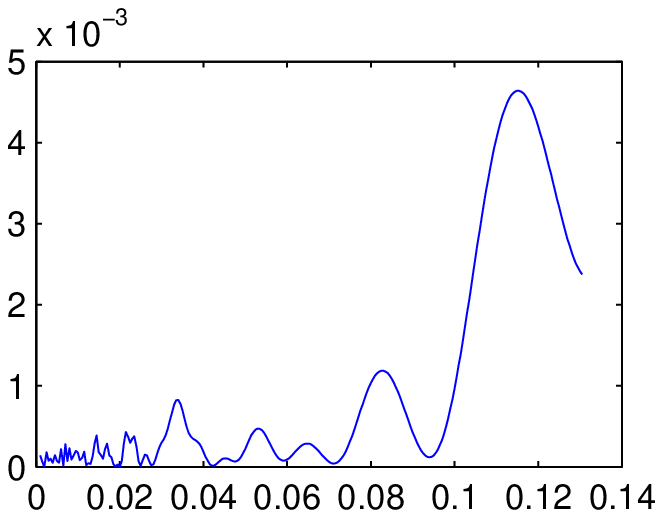}
  }
  \caption{Spectrum $K_1(T)$ of the quasar distribution according
           to the 2dF catalogue.
           Absciss corrsponds to a corresponding period $T$ over
           coordinate~$x$ from (\ref{x_variable}).
           Rayleigh criterion (\ref{Rayleigh_criterion}) used.}
  \label{2dFSpectrum1}
\end{figure}

     The result of application of criteria
(\ref{Rayleigh_criterion}) -- (\ref{Fourth_criterion}) to the SDSS
catalog in the interval $0 < T < 0.14$ is presented in
Fig.~\ref{SDSSSpectrum1}. As it follows from the plots, all the
criteria yield the similar results with diffrent confidence
degree, therefore we use below the Rayleigh criterion only. The
spectrum of the 2dF catalogue shown in Fig.~\ref{2dFSpectrum1}.

     Indeed, in the SDSS spectrum there are maxima, mentioned by
some authors, near the values $T = \Delta x=0.063$ and $0.11$,
though at a low confidence level. Except that there is one more,
even higher maximum at $T = 0.035$. In the 2dF spectrum one can
see the maxima at $T\approx 0.034, \: 0.083$ and $0.11$, but the
maximum near 0.064 is very low. Using approach of such a kind we
cannot define exact values of the confidence levels, because for
this purpose we have to simulate a stochastic sample with the same
parameters as the one in Fig.~\ref{QSODistrSDSS7}. This approach,
however, allows us to detect position of the periodic component
with highest possible precision. Note that the values of the
periodicity criterion in this case are very small. It is clear
enough, because we try to extract a weak periodic component
against a background of a very strong continuous signal, similar
to investigating of behaviour of sea waves when measuring depth of
an ocean.

\begin{figure}
  \centerline{
  \includegraphics[width=12cm]{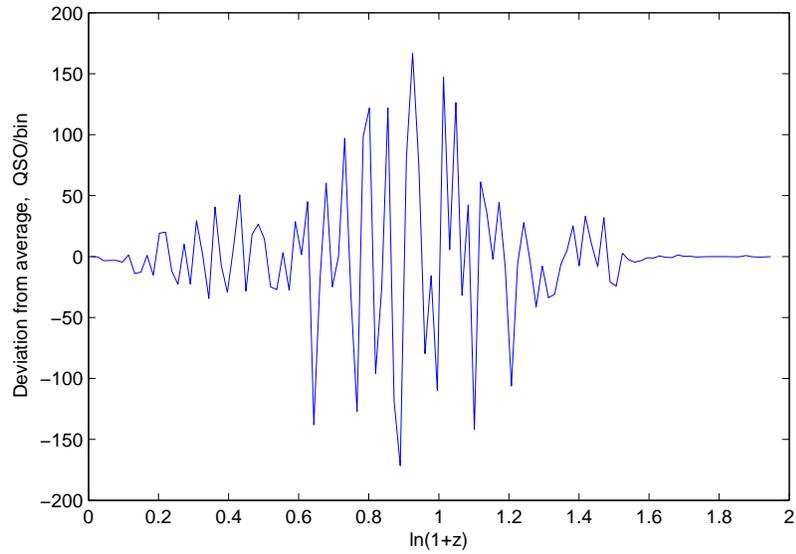}
  }
  \caption{Quasar distribution over redshift in the SDSS catalogue
           after subrtacion the background component.}
  \label{QSODistrSDSS7_2}
\end{figure}

\begin{figure}
  \centerline{
  \includegraphics[width=12cm]{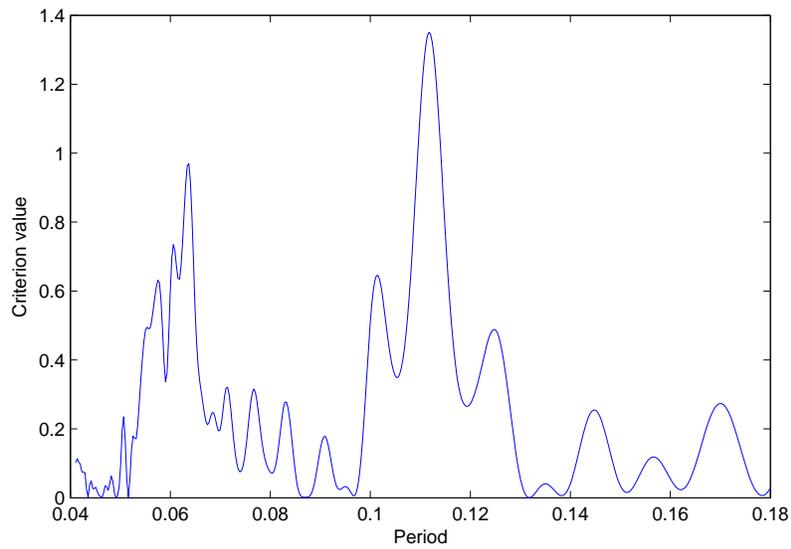}
  }
  \caption{Quasar distribution spectrum $K_1(T)$ in the SDSS
           catalogue after subtaction the background
           component.}
  \label{Spectra_2}
\end{figure}

     To increase the extraction reliability one should use
another way. Namely: one should cut the background component and
consider only the oscillations of $\Delta N_q(x)$ with respect to
a background. One can do it, subdividing the sample in narrow
$z$-intervals and analysing the numbre of quasars in these narrow
intervals, i.e. essentially, averaging the distribution inside
each interval.  The distributions in Fig.~~\ref{QSODistrSDSS7} and
~\ref{QSODistr2dF} are prepared using this technique.  As a
''background'' we consider the mean value of 5 neighbouring
intervals, where the interval of interest is in the middle. In all
cases we consider the absciss as the middle of the appropriate
interval. The result of application of this procedure to the SDSS
catalogue are presented in Fig.~\ref{QSODistrSDSS7_2} (for the 2dF
catalogue the result looks the same). If the periodic component in
the distribution does exist it should be distinguished in the plot
even by naked eye. However the plot in Fig.~\ref{QSODistrSDSS7_2}
hardly looks like a periodic one and rather looks like a noise. It
confirms the spectrum of the plot, calculated according to
Rayleigh crinterion and shown in Fig.~\ref{Spectra_2}. As it is
known \citep{gurin_1992} the mean value of the Rayleigh criterion
for the stochastic data equals to 2 and the mean standard
deviation is also 2. Indeed, we can reveal in Fig.~\ref{Spectra_2}
two maxima at $x_* = 0.063$ and 0.111, however the confidence
level does not seem to be high. One can only mark that these
components are slightly stand out against their neighbours, but
the reliable detection of the periodicity cannot be confirmed.
Thus, according to the available data we cannot draw a conclusion
that the periodic component in $\ln(1+z)$ coordinate presents in
the quasar distribution over redshift.

\begin{figure}
  \centerline{
  \includegraphics[width=12cm]{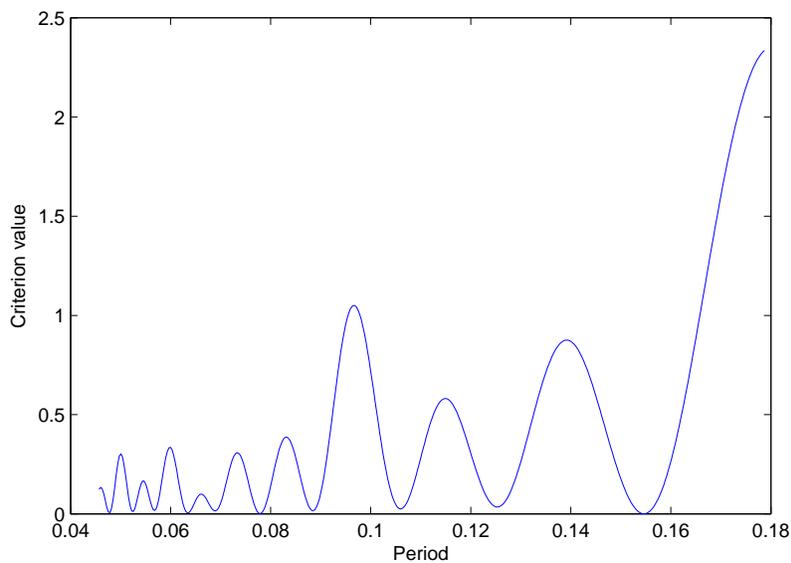}
  }
  \caption{Quasar distribution spectrum $K_1(T)$ according to the
           SDSS catalogue. The absciss axis corresponds to the
           period, espressed in the units of geodesic cosmological
           distance $R(z)$ from~(\ref{R_cosmological}).}
  \label{Spectra_4}
\end{figure}

\begin{figure}
  \centerline{
  \includegraphics[width=10cm]{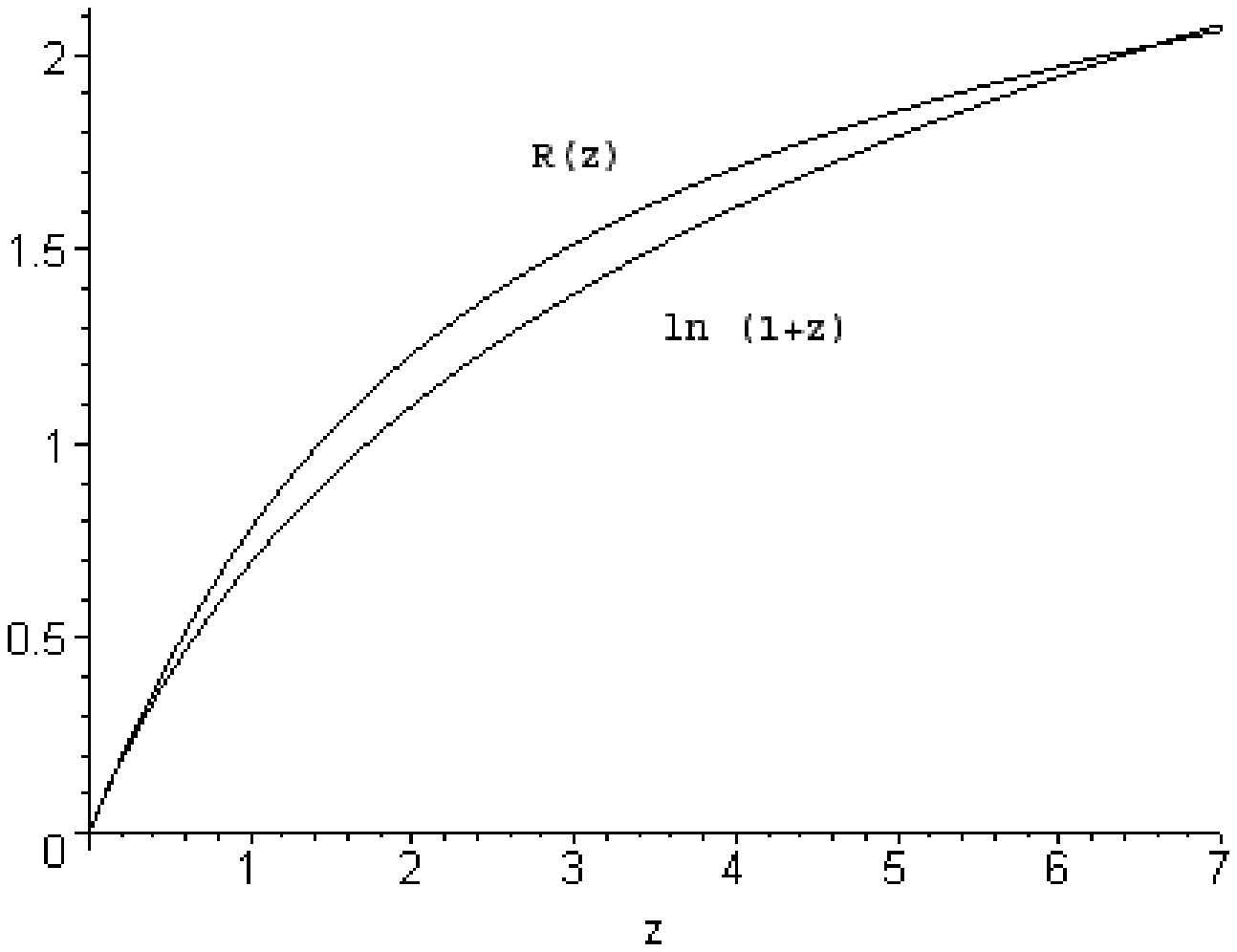}
  }
  \caption{The plots of functions $R(z)$ from (\ref{R_cosmological})
           for $\Omega_m = 0.28$, $\Omega_\Lambda = 0.72$ and
           $\ln(1+z)$. When $z \to\infty$ the function $R(z)$
           tends to a horizontal asymptote, but $\ln(1+z)$
           goes to infinity. The lines cross at
           $z=6.60517$.}
  \label{R_z_and_ln}
\end{figure}

     It, however, does not mean that the periodicity is absent for
other coordinate choice as well. It is possible to check the
periodic properties for other variables.  The most interesting
variable here is
\begin{equation}
    R(z) = \int\limits_0^z
            \frac{dz}
              {\sqrt{\Omega_m
                      \left(1 + z\right)^3 + \Omega_\Lambda}},
    \label{R_cosmological}
\end{equation}
which has the physical meaning of the geodesic cosmological
distance. Note that when $z\to \infty$ ~ $R(z)$ tends to a
constant value, i.e. to the horizon. According to current
measurements $\Omega_m + \Omega_\Lambda \approx 1$, $0.25 <
\Omega_m < 0.3$. The spectrum in $R(z)$-coordinate for the SDSS
catalogue after subtraction the background using the procedure
described above, and applying to the residual the Rayleigh
criterion~(\ref{Rayleigh_criterion}) is shown in
Fig.~\ref{Spectra_4}. Again, near the values $z_* = 0.06$ and~0.1
there are the maxima in the spectrum, but with a low confidence
level. Moreover, it seems that the maximum at $z_* = 0.06$ is
undistinguishable from the near maxima, and the peak at $z_*
\approx 0.1$ is only slightly higher than its neighbours. In
general the plots in Fig.~\ref{Spectra_2} and Fig.~\ref{Spectra_4}
do not differ enough from each other.

     The point here is that the functions $R(z)$ and $\ln (1+z)$
closely approximate each other. Both functions are shown in
Fig.~\ref{R_z_and_ln} and in the interval for $0 < z < 7$ they
differ not more than 10\%, i.e. exactly in the interval in which
drop the quasars in the SDSS catalogue. For large $z$ the
functions behave in different ways: when $z \to \infty$ $R(z)$
tends to a horizontal asymptote (for $\Omega_m =0.28$ and
$\Omega_\Lambda =0.72$ this value is 3.3988), while $\ln(1+z)$
tends to infinity.

     Thus, in this case we also cannot confirm  that in the
quasar distribution over redshift exists a periodic component.

\section{Discission and conclusions}

   From the analysis above one can draw a conclusion that the
reliable extraction of a regular periodic component in the the
quasar distribution over redshift is failed. However, the rippling
of the quasar density, probably, esceeds the statitical errors.

   Most likely  we deal with a cellular structure. The
individual cell walls, appearing on the line of sight make their
contribution to the numbre of quasars at fixed $z$. The quasar
$z$-distribution structure of such a kind is not already
stochastic, though it retains many properties of the random
distribution. In particular, the dispersion may appear greater
than theoretical, because an average amount of quasars inside the
''bubble''  and in its walls differs significantly from the
averaged numbre of the quasars in the unit volume.

    Except that at present time the SDSS catalogue covers less
than a quarter of the celestial sphere, i.e. the quasar
distribution over the celestial sphere in it is essentially
non-homogeneous. Further development of the catalogue should
smooth all irregularities in $z$-distribution. As a result the
spectral features in the distribution will be revealed at
gradually decreasing confidence level.

     Summarizing our discussion, one can draw a conclusion
that the periodic component in quasar $z$-distribution is absent
at high confidence level.

\section{Acknowledgements}

    One author (S.R.) is very grateful to
Prof.~E.Sta\-ro\-sten\-ko, Dr.~O.Su\-men\-ko\-va and
Dr.~R.Be\-res\-ne\-va for the possibility of fruitful working
under this problem. The autors are grateful to
Prof.~A.Doroshkevich and Dr.~V.Stro\-kov for their attention to
the problem and useful discusstions and Dr.~P.Iva\-nov for help.

    The work was supported by Russian Foundation for Basic
Resiarch, grants 09-02-12163, 11-02-00857 and Federal program
''Scientific and scientific-educational personnel of innovational
Russia'', State contract $\Pi$-1336.

\end{document}